\documentclass[preprint]{revtex4}

\usepackage{graphicx}
\usepackage{epstopdf} 
\DeclareGraphicsExtensions{.eps, .pdf}
\usepackage{amssymb, amsmath}
\usepackage{caption}
\usepackage[lofdepth,lotdepth]{subfig}

\begin{document}
\newcommand{\sech}{\mathop{\mathrm{sech}}\nolimits}
\newcommand{\csch}{\mathop{\mathrm{csch}}\nolimits}
\def\H{{\bar{H}}}
\def\E{{\bar{E}}}
\def\h{{\bar{h}}}
\def\e{{\bar{e}}}
\def\M{{\cal{M}}}
\def\ce{{\cal{E}}}
\def\ch{{\cal{H}}}
\def\R{{\cal{R}}}
\def\O{{\cal{O}}}
\def\T{{\bf{\cal{T}}}}
\def\t{\theta}
\def\a{\bar{a}}
\def\b{\bar{b}}
\def\z{\bar{z}}
\def\x{\bar{x}}
\def\y{\bar{y}}
\def\bxi{\bar{\xi}}
\def\u{\bar{u}}
\def\v{\bar{v}}
\def\w{\bar{w}}
\def\f{\bar{f}}
\def\g{\bar{g}}
\def\W{\tilde{W}}
\def\P{\bar{\Phi}}
\def\Ax{\bar{A_x}}
\def\Ay{\bar{A_y}}
\def\A{\bar{A}}
\def\p{\bar{\phi}}
\def\dP{\overline{\partial \Phi}}
\def\dAx{{{\overline{\partial A}}_x}}
\def\dAy{{{\overline{\partial A}}_y}}
\def\tx{{t_x}}
\def\ty{{t_y}}
\def\nx{{n_x}}
\def\ny{{n_y}}
\def\Q{{\bar{Q}}}

\title{Parametric Landau damping of space charge modes}
\author{Alexandru Macridin, Alexey Burov, Eric Stern, James Amundson, Panagiotis Spentzouris}
\affiliation{Fermilab, P.O. Box 500, Batavia, Illinois 60510, USA}

\begin{abstract}

Landau damping is the mechanism of plasma and beam stabilization; it arises through energy transfer from collective modes 
to the incoherent motion of resonant particles. Normally this resonance requires the  resonant particle's frequency
to  match the collective mode frequency. We have identified an important new damping mechanism, 
{\it parametric Landau damping}, which is driven by the modulation of the mode-particle interaction. This reveals new possibilities 
for stability control through manipulation of both particle and  mode-particle coupling spectra. 
We demonstrate the existence of parametric Landau damping in a simulation of
transverse coherent modes of bunched accelerator beams with space charge.

\end{abstract}

\maketitle

\section{Introduction}

Landau damping (LD)~\cite{landau} gives rise to the stabilization of
collective modes in plasma and accelerator beams. 
The damping is caused by the energy transfer from the collective mode to the particles 
in resonance with the mode. The damping rate is, therefore, determined by the number of the
particles capable of resonating with the mode. Conventionally, Landau damping requires the coherent resonance 
frequency
to lie within the incoherent spectrum, {\em i.e.}, to be located within the continuous frequency spectrum of the individual particles.
Manipulation of the incoherent spectrum is often proposed as a mean to enhance stability; for example,
the increase of the betatron tune spread  in accelerators by using nonlinear magnets~\cite{cern-091}.
In this paper we discuss a new mechanism for Landau damping which occurs when the mode-particle coupling 
has an extended frequency spectrum. This  mechanism reveals new possibilities for stability enhancement involving 
both particles' and the mode-particle coupling's spectra.

The novel Landau damping mechanism, which we call {\em parametric Landau damping}, is 
revealed by the  numerical investigation of the transverse space charge  modes in
accelerator bunched beams at the coupling resonance, {\em i.e.}, when the horizontal
and the vertical tunes are close. In contrast with the usual Landau damping mechanism, the frequency of the LD-responsible particles, 
{\em i.e.}, the particles which absorb the mode energy, have a wide spread and do not match the coherent
frequency. This happens due to the modulation of the mode-particle coupling by  particle dependent
oscillations with a wide frequency spread.

The transverse space charge modes in bunched beams, away from the coupling resonance
were found analytically in Refs~\cite{blaskiewicz_prstab_1998,burov_prstab_2009,balbekov_prstab_2009}.
Their intrinsic Landau damping in the strong space charge regime was suggested in Ref~\cite{burov_prstab_2009}. 
Predicted damping rates were confirmed by numerical 
simulations~\cite{kornilov_prstab_2010, kornilov_HB2014,macridin_prstab_2015}.

The effect of linear coupling resonance on Landau damping was addressed by Metral in~\cite{metral_thesis}. 
He showed that in the proximity of linear coupling resonance, the incoherent frequency spectrum from both 
transverse planes contributes to  Landau damping. The collective mode transfers energy to the
incoherent motions in both planes.  Metral's mechanism is  a shared  damping 
between the transverse planes;  
a damping increase  of the modes present in one plane
implies a compensating decrease in the damping of the modes present in the other plane.
In contrast, our mechanism is due to an oscillating mode-particle coupling implying additional
conditions for the resonant energy transfer. Since our mechanism does not involve sharing, it is possible  to enhance  Landau damping 
for modes present in both planes.
In our case the coupling between the transverse planes is produced by
the space charge force, thus no linear terms are present.
The main  resonance is the fourth-order Montague resonance~\cite{Montague}.
In our simulations the frequency of many LD-responsible particles  
does not match the coherent frequency.
The enhancement of the damping rate cannot be explained by 
Metral's coupling resonance mechanism which requires 
the presence of resonant particles around the coherent frequency in the plane perpendicular to the mode.

A common feature of the particles trapped in the vicinity of resonance fixed points  is the oscillation 
of their amplitudes. In the coupling resonance case the trapped particles are characterized by an oscillatory energy exchange 
between the transverse planes.  Their transverse amplitudes
are oscillating with typical trapping frequencies. Since the  mode-particle coupling
is dependent on the particles' amplitudes, it is modulated by the 
trapping frequency. The  resonance condition for Landau damping  requires the particle frequency to equal the mode
frequency shifted by the trapping frequency. Because the trapping frequencies are particle dependent, 
the frequencies of the LD-responsible particles may span a range equal to the one of the
mode-particle coupling frequency spectrum.

We employ the Synergia
accelerator modeling package~\cite{synergia,amundson_jcp_2006}  to  simulate the propagation of a 
single Gaussian beam through a linear lattice.
The modes are extracted from the transverse displacement density using the dynamic mode decomposition 
(DMD) technique~\cite{dmd1,dmd2,dmd3,dmd4}. DMD is a data driven algorithm used for modal analysis 
in both linear and nonlinear systems.

We compare the properties of the first space charge mode
away  and in the vicinity of the coupling resonance.
We find that the Landau damping is larger in the latter
case, while the frequency and the mode shape are nearly the same. By investigating the
properties of the particles exchanging energy with the mode, we conclude that the off-resonance
case well fits the conventional Landau damping scenario characterized by  LD-responsible particles
with an incoherent frequency spectrum at the coherent frequency.
Around coupling resonance the damping enhancement is caused by the parametric Landau damping mechanism, a consequence of 
the modulated coupling between the mode and the trapped particles. 
The existence of the parametric Landau damping mechanism  for the first space charge mode at coupling resonance is proven  solely by  numerical 
simulations of a bunch propagating through a lattice; no analytical model is assumed. 

The paper is organized as follows. In Sec.~\ref{sec:pld} the mechanism of
parametric Landau damping is  formally introduced. In Sec.~\ref{sec:scm} the Landau damping mechanism of
transverse space charge modes at coupling resonance is discussed. The details of the simulations are 
described in Sec.~\ref{sec:synergia}. The results of the simulations
are presented in Sec.~\ref{sec:results}, followed by  discussions in Section~\ref{sec:discuss}.
Summary and conclusions are given in Section~\ref{sec:conclusions}.
In Appendix~\ref{app:tsham} the calculation of tune shift and mode-particle coupling is discussed.

\section{Parametric Landau damping}
\label{sec:pld}

The Landau damping mechanism results from the interaction of the collective mode with the individual particles.
Using the simple harmonic oscillation approximation~\cite{Chao_book}, the equation of motion for the particle $i$ interacting with the mode $\x$
can be written as  
\begin{eqnarray}
\label{eq:gmotion}
\ddot{x}_i+\omega_i^2  x_i =-K_i\x(t), 
\end{eqnarray}
\noindent where $x_i$ represents the particle displacement, $\omega_i$ the particle frequency, $K_i$ 
the mode-particle coupling and $\x(t)$ the collective mode.

In systems with conventional Landau damping,  $K$ is either time independent or its oscillation
frequency is particle independent.
The resonance condition is $\omega_i = \omega_c$, where $\omega_c$
is the $\x(t)$ frequency, {\em i.e.}, $\x(t) \propto \exp(-i \omega_c t)$.
The damping rate is proportional to the spectral density at the resonant frequency, 
\begin{eqnarray}
\label{eq:dampingLandau damping}
\lambda \propto \rho(\omega_c)=\sum_i \delta(\omega_i-\omega_c).
\end{eqnarray}

Nevertheless, as in the case addressed in this paper,
it may happen that the mode-particle coupling is characterized by a frequency spectrum, {\em i.e.},
$K_i(t) \propto \exp(-i \mu_i t)$ and $\mu_i$ is particle dependent. The resonance condition
in this case is $\omega_i=\omega_c \pm \mu_i$. 
The damping rate is proportional to the number of particles which fulfill the resonance condition,
\begin{eqnarray}
\label{eq:dampingPLandau damping}
\lambda \propto h(\omega_c)=\sum_i \delta(\omega_c-\omega_i \pm \mu_i).
\end{eqnarray}
\noindent In this case the damping 
is determined by the interplay of both particles and mode-particle coupling spectra. We 
call this  mechanism parametric Landau damping.

\section{Landau damping of transverse space charge modes}
\label{sec:scm}

For transverse space charge modes the equation of motion for the particle $i$ can be written as 
(see Appendix~\ref{app:tsham}, Eq.(\ref{eq:cmk}), \cite{mohl_schonauer,blaskiewicz_prstab_1998, burov_prstab_2009, balbekov_prstab_2009})
\begin{eqnarray}
\label{eq:motion}
\ddot{x}_i+\omega_0^2 \left(Q_{0x}-\delta Q_{x}(z_i,J_{xi},J_{yi})\right)^2 x_i =-K(z_i,J_{xi},J_{yi}) \bar{x}(t,z_i),
\end{eqnarray}
\noindent where $\omega_0$ is the angular revolution frequency and $Q_{0x}$ is the bare betatron tune.
The tune shift $\delta Q_{x}(z_i,J_{xi},J_{yi}) \ll Q_{0x}$ and the mode-particle coupling $K(z_i,J_{xi},J_{yi})$ are proportional 
to the longitudinal 
density and are dependent on the particle transverse actions, $J_{xi}$ and $J_{yi}$, defined as
\begin{eqnarray}
 \label{eq:Jx}
 J_{xi}=\frac{x_i^2+(\alpha_x x_i+ \beta_x x'_i)^2}{2 \beta_x},~~ J_{yi}=\frac{y_i^2+(\alpha_y y_i+ \beta_y y'_i)^2}{2 \beta_y},
\end{eqnarray}
\noindent where $\alpha_x$, $\beta_x$, $\alpha_y$ and $\beta_y$ are the lattice Twiss parameters.

The coherent driving term $\bar{x}(t,z_i$) in Eq.(\ref{eq:motion}) is  the average displacement 
at the location where the particle is currently located and can be written as $\bar{x}(t, z_i)=e^{-i \omega_0 \nu t} \bar{x}[z_i(t)]$, 
where
\begin{eqnarray}
\label{eq:barx}
\bar{x}(z)=\frac{\int dx dx' dy dy' du x \rho(x,x',y,y',z,u)}{ \rho(z)},
\end{eqnarray}
\noindent and $\nu$ being the mode tune~\cite{burov_prstab_2009}. 
Here $z$ is the longitudinal position relative to the reference particle, $u=\frac{\delta p}{p}$ is the relative momentum spread,
$\rho(x,x',y,y',z,u)$ is the density in $6D$ phase space and $\rho(z)=\int dx dx' dy dy' du \rho(x,x',y,y',z,u) $ 
is the longitudinal density.
Unlike the situation described by Eq.(\ref{eq:gmotion}), in Eq.(\ref{eq:motion}) the uncoupled particle motion, the mode particle coupling 
and $\x$ are not characterized by a single frequency. 

In order to address the Landau damping mechanism in the following part of this section 
we will investigate the conditions for resonant coupling between the 
particles and the collective mode. The explicit dependence of the tune shift $\delta Q_{x}$ and the mode-particle coupling $K$
on the particle's action coordinates is not important for this analysis.

In the off-resonance case, to a good approximation
the particle transverse actions $J_{xi}$ and $J_{yi}$  are constants of motion and the time dependence of
the tune shift $\delta Q_{x}(z_i,J_{xi},J_{yi})$ is a consequence of the synchrotron motion given by
$z_i(t) =a_i \cos(\omega_0 Q_s t +\varphi_i)$.  Here $a_i$ is the  particle's longitudinal amplitude,
$Q_s$ is the synchrotron tune and $\varphi_i$ is the particle's  phase at $t=0$.  The tune shift $\delta Q_{x}(z_i,J_{xi}, J_{yi})$ 
for a Gaussian beam with the longitudinal density $\rho(z)=\exp(-z^2/2\sigma_z)$ can be expanded  as
\begin{eqnarray}
\label{eq:Qzi}
 \delta Q_{x}[z_i(t),J_{xi}, J_{yi}] & = &\delta Q_{x}[0,J_{xi}, J_{yi}] \exp \left( \frac{-z^2_i(t)}{2 \sigma_z^2} \right) \\ \nonumber
&=& \delta Q_{x}[0,J_{xi}, J_{yi}] \exp \left(  \frac{-(a_i \cos (\omega_0 Q_s t +\varphi_{i}))^2}{2 \sigma_z^2} \right) \\ \nonumber
&=&C_{0i}+\sum_{m=1}^{\infty} 2 C_{mi} \cos \left[2 m (\omega_0 Q_s t+\varphi_{i})\right],
\end{eqnarray}
\noindent  where
\begin{eqnarray}
\label{eq:cm}
C_{mi}=\delta Q_{x}[0,J_{xi}, J_{yi}](-1)^m \exp \left( {-\frac{a_i^2}{4 \sigma_z^2}} \right) I_m(\frac{a_i^2}{4 \sigma_z^2}),
\end{eqnarray}
\noindent and $I_m$ are the modified Bessel functions of the first kind. The tune shift contains a constant term
which depends on the particle's longitudinal 
amplitude $a_i$ and higher harmonics of $2Q_s$ terms. Since the mode-particle coupling $K(z_i,J_{xi}, J_{yi})$ 
is also proportional to the charge density $\rho(z)$, an analogous  expansion is valid for the mode-particle coupling term
\begin{eqnarray}
\label{eq:Ki}
 K[z_i(t),J_{xi}, J_{yi}]  =K_{0i}+\sum_{m=1}^{\infty} 2 K_{mi} \cos \left[2 m (\omega_0 Q_s t+\varphi_{i})\right].
\end{eqnarray}

The mode $\x(t,z_i)$ frequency as seen by the particle $i$, is also influenced by the synchrotron motion.
For that reason, the main tune of the first space charge mode is $\nu-Q_s$ and not $\nu$.
This can be understood by considering the approximation 
$\bar{x}[z] \approx \sin [\pi z/4 \sigma_z]$~\cite{blaskiewicz_prstab_1998,kornilov_prstab_2010} for the first space charge mode.
The particle $i$ sees the mode as
\begin{eqnarray}
\label{eq:bar_x1z}
\bar{x}[t,z_i(t)] &\approx& e^{-i \omega_0 \nu t} \sin (\frac{\pi a_i \cos (\omega_0Q_s t +\varphi_{i})}{4 \sigma_z}) \\ \nonumber
&= &e^{-i \omega_0 \nu t} \left[ J_1(\frac{\pi a_i}{4 \sigma_z}) \cos (\omega_0Q_s t+\varphi_{i})+higher~harmonics \right].
\end{eqnarray}
\noindent Here $J_1$ is the Bessel function of the first kind and $higher~harmonics$ represent $(2n+1)Q_s$  higher harmonic terms  
proportional to higher order Bessel functions. They are smaller in magnitude compared to the $J_1$ term.

From  Eqs.(\ref{eq:motion}), (\ref{eq:Qzi}), (\ref{eq:Ki}) and (\ref{eq:bar_x1z}) one can conclude that, in the off-resonance case,
the main energy resonant exchange specific to the Landau damping mechanism between the first space charge mode 
and the particles takes place at the tune
$\nu-Q_s$. Resonant exchanges at tunes distanced from $\nu-Q_s$  by  harmonics  of $2Q_s$ are also present, but
we find that they  play only a minor role in the Landau damping mechanism, as 
discussed in Sec.~\ref{sec:results}.

In the off-resonance case the Landau damping is conventional, since
the  oscillation frequencies of the mode-particle coupling are particle independent ($K$ contains
a constant term and harmonics of $2Q_s$, see Eq.(\ref{eq:Ki})).
The situation is different at the coupling resonance. Unlike the off-resonance case, 
in the proximity of the coupling resonance $J_{xi}$ and $J_{yi}$ are not constants of motion.
In fact the sum $J_{si}=J_{xi}+J_{yi}$ is a constant of motion, 
while the difference $J_{di}=J_{xi}-J_{yi}$ oscillates around the stable point with the trapping frequency
$\omega_{ti}=\omega_0 Q_{ti}$ (see Eq.(\ref{eq:app_wtr})). The essential feature for the parametric Landau damping mechanism is that the particle dependent $Q_{ti}$  
modulates the mode-particle coupling. This becomes  evident when the mode-particle coupling is written as (see Eq.(\ref{eq:app_gpmc}))
\begin{eqnarray}
\label{eq:K_AB}
K(z_i,J_{xi},J_{yi})= A(z_i,J_{si})+B(z_i,J_{si},J_{di}) J_{di}.
\end{eqnarray}
The terms $A(z_i,J_{si})$ and $B(z_i,J_{si},J_{di})$ as well as the tune shift $\delta Q_x(z_i,J_{si},J_{di})$  depend on the beam shape
and can be approximated analytically for certain cases. We discuss the calculation
of the tune shift and the mode-particle coupling in  Appendix~\ref{app:tsham}. 
In the Appendix~\ref{app:tcr}, to give an example,
we calculate the particles' dynamics in the proximity of the coupling resonance stable point for a 2 dimensional (2D) Hamiltonian
with a fourth order, rotationally symmetric transverse coupling term.  Equation (\ref{eq:app_gpmc}) gives  the mode-particle
coupling in our example. A more general Hamiltonian describing the dynamics at coupling
resonance, but without the presence of  collective  modes,  
was considered  by Montague~\cite{Montague}. 

The equation of motion in the proximity of the coupling resonance is 
\begin{eqnarray}
\label{eq:motionJd}
\ddot{x}_i+\omega_0^2 Q_{x}^2(z_i,J_{si},J_{di}) x_i=-A(z_i,J_{si})\bar{x}(t,z_i)-B(z_i,J_{si},J_{di}) J_{di}\bar{x}(t,z_i),
\end{eqnarray}
\noindent where $Q_x(z_i,J_{si},J_{di})=Q_{0x}-\delta Q_x(z_i,J_{si},J_{di})$.
Since $J_{si}$ is a constant of motion, 
the coupling term $A\x$ is conventional. The term $BJ_d\x$ yields parametric Landau damping because
it is modulated by oscillations with $\omega_0 Q_{ti}$ frequency. 
While in the former case the resonance condition
is  $\Q_{xi} = \nu-Q_s$, the  $BJ_d\bar{x}$ term implies mode-resonant  particles when $\Q_{xi}  = \nu-Q_s - Q_{ti}$.
Here $\Q_{xi}$ is the particle $i$ main tune defined as the tune of the largest peak in the Fourier spectrum of $x_i(t)$.
In principle the  dependence of $B(z,J_s,J_d)$ on $J_d$  yields 
resonant energy exchanges at frequencies spaced by harmonics of $Q_t$ from the main parametric resonant condition, 
{i.e. \em}, $\Q_{xi}  = \nu-Q_s - Q_{ti}+n Q_{ti}$,
but in our simulations we find these processes  not to be significant.

The oscillations of $J_d$ 
contribute not only to the parametric Landau damping but to the
conventional one as well. The dependence of  $Q_x(z_i,J_{si},J_{di})$ in Eq.(\ref{eq:motionJd}) 
on $J_{di}$ yields satellite features
spaced by harmonics of $Q_{ti}$ in the incoherent spectrum. These satellites
are resonant via $A\bar{x}$ coupling when their tune is equal to $\nu-Q_s$.

\section{Simulations}
\label{sec:synergia}

The simulations are  done by employing Synergia~\cite{synergia}, a particle tracking code
for beam dynamics in accelerators. The space charge effects are implemented in Synergia using the 
second order split-operator method~\cite{yoshida}.  At every step, the electric field is calculated by solving 
 the 3D Poisson equation with open boundary conditions numerically as described in~\cite{hockney}.

The bunch is initially excited in the horizontal plane with the first space charge harmonic function. 
Space charge harmonics are the space charge modes of Gaussian beams in the strong space charge limit and were calculated analytically in 
Refs.~\cite{burov_prstab_2009,balbekov_prstab_2009}. 
The excitation amplitude is small enough to ensure linear damping regime and not to affect the particles' tune spectrum.
The transverse displacement density, 
\begin{eqnarray}
\label{eq:xdipole_def}
X(z,u,s)=\frac{\int dx dx' dy dy' x \rho(x,x',y,y',z,u,s)}{ \rho(z,u,s)},
\end{eqnarray} 
\noindent is calculated at every turn. Here $s$ is the distance along the reference trajectory,
and 
\begin{eqnarray}
\label{eq:rhozu}
\rho(z,u,s)=\int dx dx' dy dy' \rho(x,x',y,y',z,u,s)
\end{eqnarray} 
\noindent  is the density in the longitudinal phase space.
The modes' shape, tune and damping are extracted from $X(z,u,s)$ using the DMD technique. 
DMD~\cite{dmd1,dmd2,dmd3,dmd4} has been 
used  for mode analysis in many fields such as fluid 
mechanics~\cite{dmd_examples1}, neuroscience~\cite{dmd_examples2},  and video streaming and pattern 
recognition ~\cite{dmd_examples3}.
An important advantage of this method is 
the direct calculation of mode dynamics, including shape, frequency, and growth/damping rates.
Application of Synergia and DMD to beam dynamics is described in detail in~\cite{macridin_prstab_2015}.

A lattice with the length  $200~\text{m}$ made by $10$ identical OFORODO 
(drift - focusing quad - drift - rf cavity - drift - defocusing quad - drift) 
cells is chosen.
The difference between the actual phase advance and the smooth 
approximation phase advance nowhere exceeds $2\%$  of the phase advance per cell. 
For the off-resonance case we take the bare betatron tune 
difference $Q_{0x}-Q_{0y} > \delta Q_{sc~max}$ while at the coupling resonance  $Q_{0x}=Q_{0y}$.
$\delta Q_{sc~max}$ is the space charge tune shift at the center of the bunch.
The tunes in our simulations have values in the range typical for real circular accelerators,
with $Q_s \ll Q_{0x}, Q_{0y}$. The majority of the simulations are done with
$Q_{0x}=2.322$ and $Q_s=0.01$. We checked the robustness of our results for    
other values of the tunes, such as $Q_{0x}=2.443$ and $Q_s=0.005$.
A proton beam with the energy corresponding to the relativistic factor $\gamma=1.6$ is chosen.
The chromaticity is zero.
$10^8$  macroparticles per bunch are used for the simulations.
The beam distribution is longitudinally and transversely Gaussian with equal vertical and horizontal 
emittances, $\epsilon^{rms}_x=\epsilon^{rms}_y=1~\text{mm*mrad}$.
The  space charge parameter is defined as $q=\frac{\delta Q_{sc~max}}{Q_s}$.

\section{Results}
\label{sec:results}

While the formalism described in Sections~\ref{sec:pld} and~\ref{sec:scm} and Appendix~\ref{app:tsham}
is useful for understanding the damping mechanism and interpreting the simulations,  
the results presented in this section are based only on the tracking simulations
of a Gaussian beam  through an  OFORODO lattice, as described in Sec~\ref{sec:synergia}.

\begin{figure}

\begin{center}
\includegraphics*[width=5.5in]{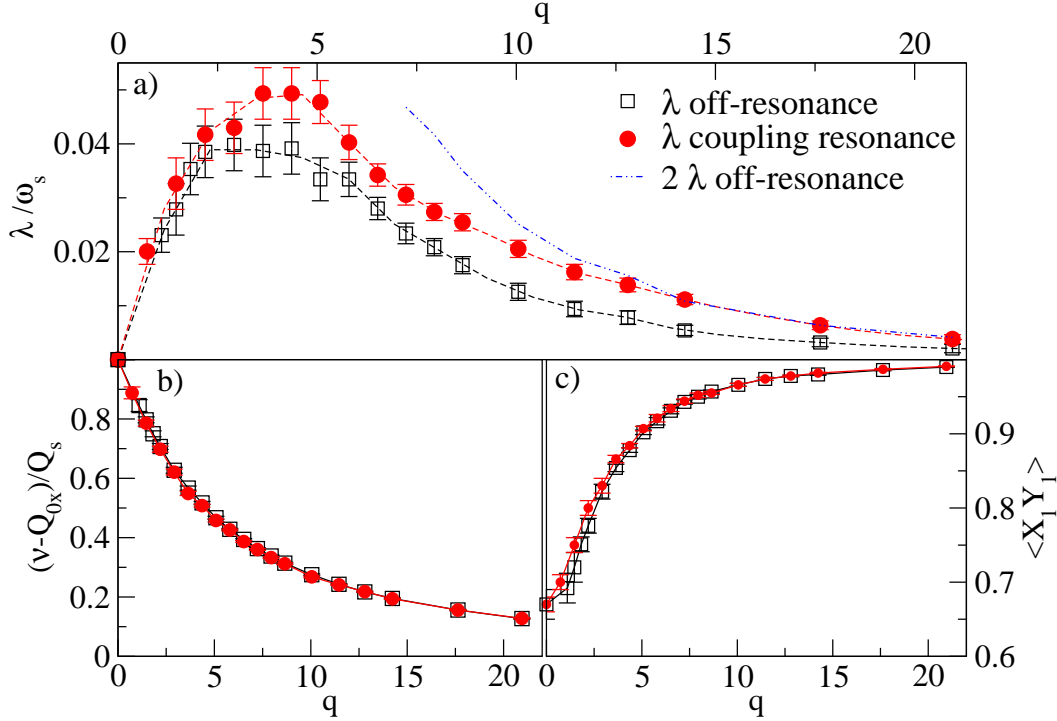}
\caption{Comparison between the off-resonance and the coupling resonance cases for the first space charge mode:
(a) Landau damping rate $\frac{\lambda}{\omega_s}$ versus space charge parameter $q$, where $\omega_s=\omega_0 Q_s$ is 
the synchrotron frequency. 
For intermediate and strong space charge 
the damping is larger at the coupling resonance.  (b) The mode tune $\nu$ at off-resonance and at coupling resonance are nearly the same.
(c) The spatial overlap, Eq.(\ref{eq:mode1_overlap}), of the DMD extracted mode $X_1(z,u)$ with the space charge harmonic 
function $Y_1(z)$. The off-resonance and the coupling resonance mode shapes are nearly the same.
}
\label{fig:mode1_comp}
\end{center}
\end{figure}

The properties of the first space charge mode off-resonance and at coupling resonance
are compared in Fig.~\ref{fig:mode1_comp}. For intermediate and large space charge,
$q\gtrsim4$, the  damping at  coupling resonance is larger, as shown in Fig.~\ref{fig:mode1_comp}(a).
In the strong space charge regime, $10\lesssim q \lesssim 20$, the damping at coupling resonance
is larger by approximately a factor of $2$.
The relative enhancement increases slowly with $q$. 
However the precision of the simulation at large $q$ is limited by the small value of the 
Landau damping, which becomes of the order of the error bars.
The mode tune measured relative to the bare betatron tune, $\nu-Q_{0x}$, is nearly the same for both cases, see Fig.~\ref{fig:mode1_comp}(b).
The difference between the mode spatial shape in the two cases is  also insignificant,
as illustrated in Fig.~\ref{fig:mode1_comp}(c) where the spatial overlap of the mode, $X_1(z,u)$, with the first space charge harmonic $Y_1(z)$ 
($Y_1(z)$ is calculated 
in~\cite{burov_prstab_2009}),
\begin{eqnarray}
\label{eq:mode1_overlap}
<X_1 Y_1> = \int X_1(z,u)Y_1(z) \rho(z,u) dz du,
\end{eqnarray}
\noindent is plotted.

\begin{figure}
\includegraphics*[width=5.5in]{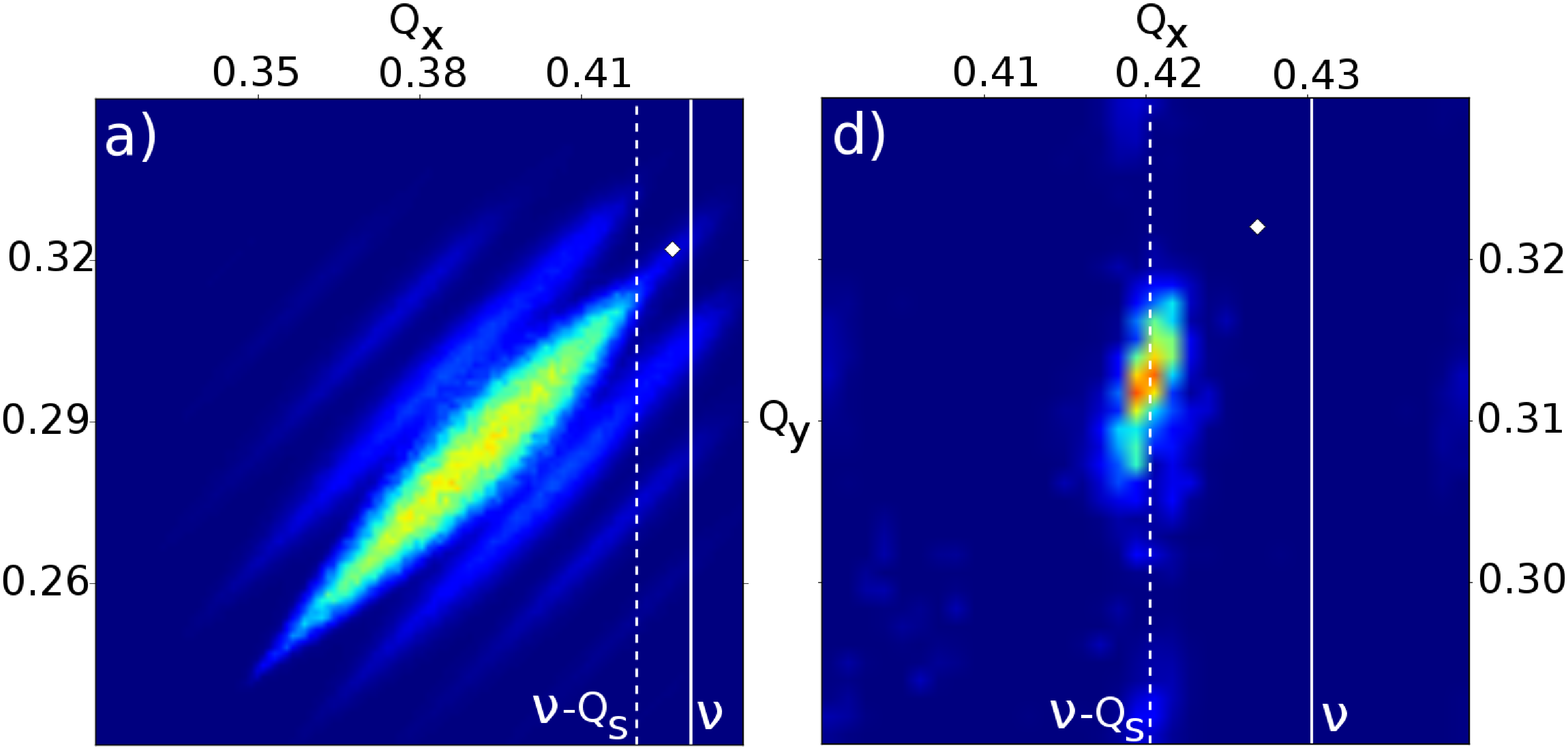}
\includegraphics*[width=5.5in]{./Jvsturn_d1.eps}
\caption{(a) Bunch tune footprint at off-resonance for $q=7.94$. The white dot corresponds to
the bare betatron tunes. (b) $\sum_{i \in S} \Delta J_x=\sum_{i \in S}\left( J_{xi}-J_{xi~initial}\right)$ and 
$\sum_{i \in S} \Delta J_{yi}=\sum_{i \in S} \left( J_{yi}-J_{yi~initial}\right)$ 
of the $0.05\%$ and $0.2\%$ largest increasing energy particles versus turn number, normalized by the product of emittance 
and the number of the particles in the sum. (c) The same as (b) but for
the largest decreasing energy particles. (d) Tune footprint for the $0.5\%$ largest changing energy (increase and decrease) particles. 
The tunes are in the proximity of the  coherent tune $Q_x=\nu-Q_s$.
The color dimension scale in (a) and (d) differs by one order of magnitude.}
\label{fig:d1}
\end{figure}

The off-resonance Landau damping mechanism can be understood within
the typical paradigm. In Fig.~\ref{fig:d1} (a) we plot the beam 2D tune footprint $\rho(Q_x,Q_y)$
defined as
\begin{eqnarray}
\label{eq:rhoi}
\rho(Q_x,Q_y)= \sum_i |\tilde{x}_i(Q_x)|^2 |\tilde{y}_i(Q_y)|^2,
\end{eqnarray}
\noindent where
\begin{eqnarray}
\label{eq:ftx}
\tilde{x}_i(Q_x) = \frac{1}{C_{xi}} \int x_i(s) e^{i \omega_0 Q_x s} ds
\end{eqnarray}
\noindent is the  Fourier transform of the particle $i$ horizontal displacement $x_i(s)$. An analogous definition 
is used for $\tilde{y}_i(Q_y)$.
Since the Landau damping is determined by the tune 
density and is insensitive to the particles' amplitude {\em per se}, the constant $C_{xi}$ in Eq.(\ref{eq:ftx}) is used to
normalize the spectral weight of each particle, Eq.(\ref{eq:rhoi}), to one. As a consequence, the integrated $\rho(Q_x,Q_y)$ 
over the horizontal and the vertical frequencies is equal to the number of particles in the bunch. 
The space charge force shifts the particles' tunes to lower values. 
The tune depression is maximal at the bunch center,
while the particles in the distribution tails have a much smaller tune shift.
The satellite lines separated by $2 Q_s$  are a consequence of the modulation of the tune shift with 
the particle's longitudinal position (as Eq.(\ref{eq:Qzi}) predicts). The  particles directly responsible for the Landau damping
are the ones which resonantly exchange energy with the mode.
To select the LD-responsible particles we look for those having the largest change
in their energy  between the end and the beginning of  the simulation.
In Fig.~\ref{fig:d1}(b)  we plot 
\begin{eqnarray}
\label{eq:sumJx}
\sum_{i \in S} \Delta J_x= \sum_{i\in S}\left( J_{xi}-J_{xi~initial}\right)~~\text{and}~~\sum_{i \in S} \Delta J_y= \sum_{i\in S}\left( J_{yi}-J_{yi~initial}\right)
\end{eqnarray}
\noindent where $J_{xi~initial}$ and $J_{yi~initial}$ represent the particle $i$ actions at the beginning of the simulation.
The notation $S$ represents the subset of the $0.05\%$ (black) or the $0.2\%$ (red and blue) 
largest energy increase  particles. A similar plot is shown in Fig.~\ref{fig:d1}(c), but for the
largest energy decrease particles. The values $0.05\%$ and $0.2\%$ are arbitrary chosen  for the purpose of illustrating the properties 
of the LD-responsible particles. In the linear Landau damping theory, the energy of 
the LD-responsible particles increases linearly
in a time interval $\delta t \approx 1/|\delta \omega|$, where $\delta \omega$ is
the frequency difference between the particle and the mode. As  seen from Fig.~\ref{fig:d1}(b), 
the time where $\sum \Delta J_x$  is increasing linearly is larger when the number of the
particles in the summation is smaller, since a larger number in the summation implies
particles with larger $|\delta \omega|$.
 Note that the chosen particles increase or decrease their action only in the horizontal plane,
{\em i.e.}, the plane where the mode is present. The 2D tune footprint of the $0.5\%$  largest-energy changing
particles (both increasing and decreasing) is shown in  Fig.~\ref{fig:d1}(d). As expected, since these particles are mode-resonant,
their horizontal tune is in the vicinity of $\nu-Q_s$. Notice that higher $\nu-(2n+1)Q_s$ harmonics spectral features 
do not appear to be significant in the tune footprint of the LD-responsible particles.

\begin{figure}
\includegraphics*[width=5.5in]{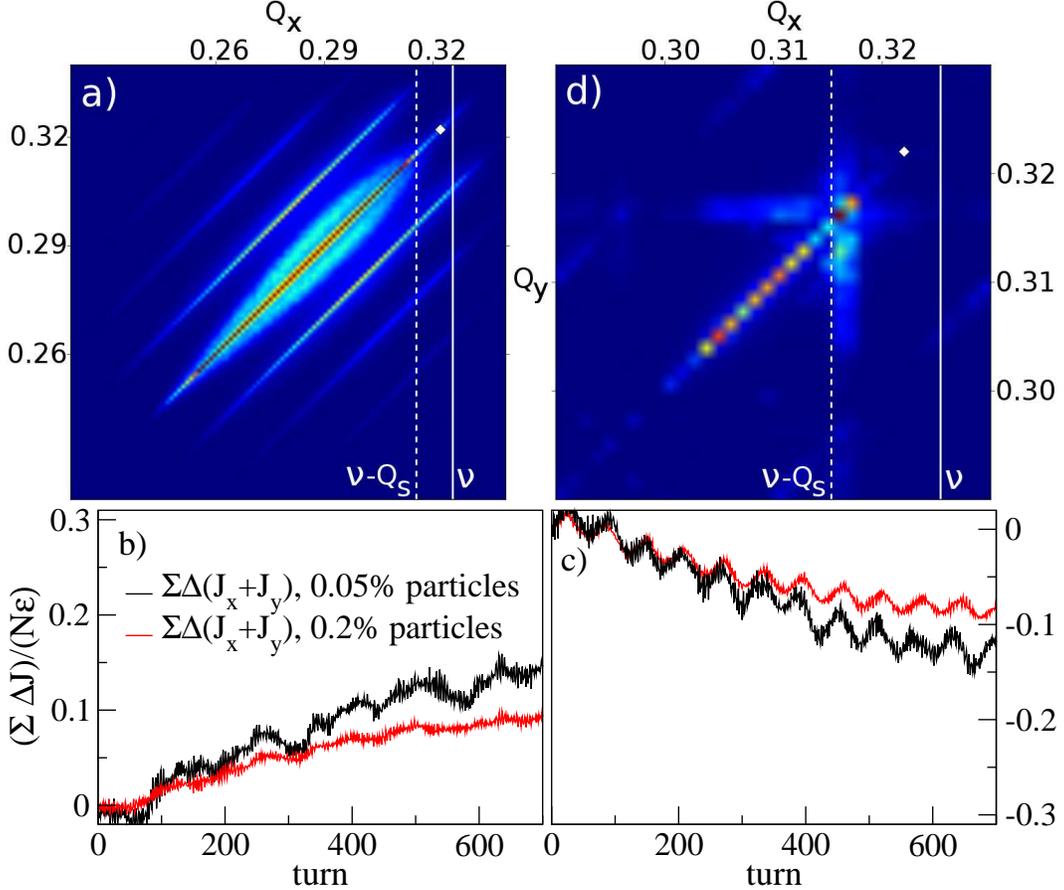}
\includegraphics*[width=5.5in]{./Jvsturn_d0.eps}
\caption{(a) Bunch tune footprint at coupling resonance for $q=7.94$. The white dot corresponds to
the bare betatron tunes. (b) $\sum_{i \in S} \Delta J_{si}=\sum_{i \in S} \left(J_{si}-J_{si~initial} \right)$ 
of the $0.05\%$ and $0.2\%$ largest increasing energy  particles versus turn number, normalized by the product of emittance and the number of the particles in the sum. 
(c) The same as (b) but for
the largest decreasing energy  particles. 
(d) Tune footprint for the $0.5\%$ largest changing energy (increase and decrease)
particles. Large part of the spectral weight is along the resonance line $2Q_x-2Q_y=0$,
with the horizontal tune well below $\nu-Q_s$.
The color dimension scale in (a) and (d) differs by one order of magnitude.
}
\label{fig:d0}
\end{figure}

The spectral properties of the LD-responsible particles at the coupling resonance 
do not fit the typical Landau damping paradigm. The beam 2D tune
footprint in Fig.~\ref{fig:d0}(a) displays enhanced spectral weight along the coupling 
resonance line $2Q_{x}-2Q_{y}=0$, consequence of resonance trapping 
(in agreement with Eqs.(\ref{eq:app_xtune_shr}) and (\ref{eq:app_ytune_shr})).
Satellite lines corresponding to the $2Q_s$ harmonics can be observed too.
We use the same largest energy change criterion to select
the LD-responsible particles. Unlike the off-resonance case, the horizontal and vertical
actions exhibit non-monotonic change with turn number, since in the proximity of coupling resonance
their magnitude oscillates between the planes. However, the transverse action sum $J_s$ of the LD-responsible particles
displays a monotonic increase (decrease),  as shown in  Fig.~\ref{fig:d0}(b)(Fig.~\ref{fig:d0}(c)).
The interesting fact which points to a parametric damping mechanism
is that the tune of most of these large energy changing particles is not in the 
vicinity of $\nu-Q_s$ as one would expect for  LD-responsible particles. As shown in
Fig.~\ref{fig:d0}(d), there is a large spectral weight on the coupling resonance line
which extends well below  $Q_x=\nu-Q_s$.

\begin{figure}
\begin{center}
\subfloat{\includegraphics*[width=2.75in]{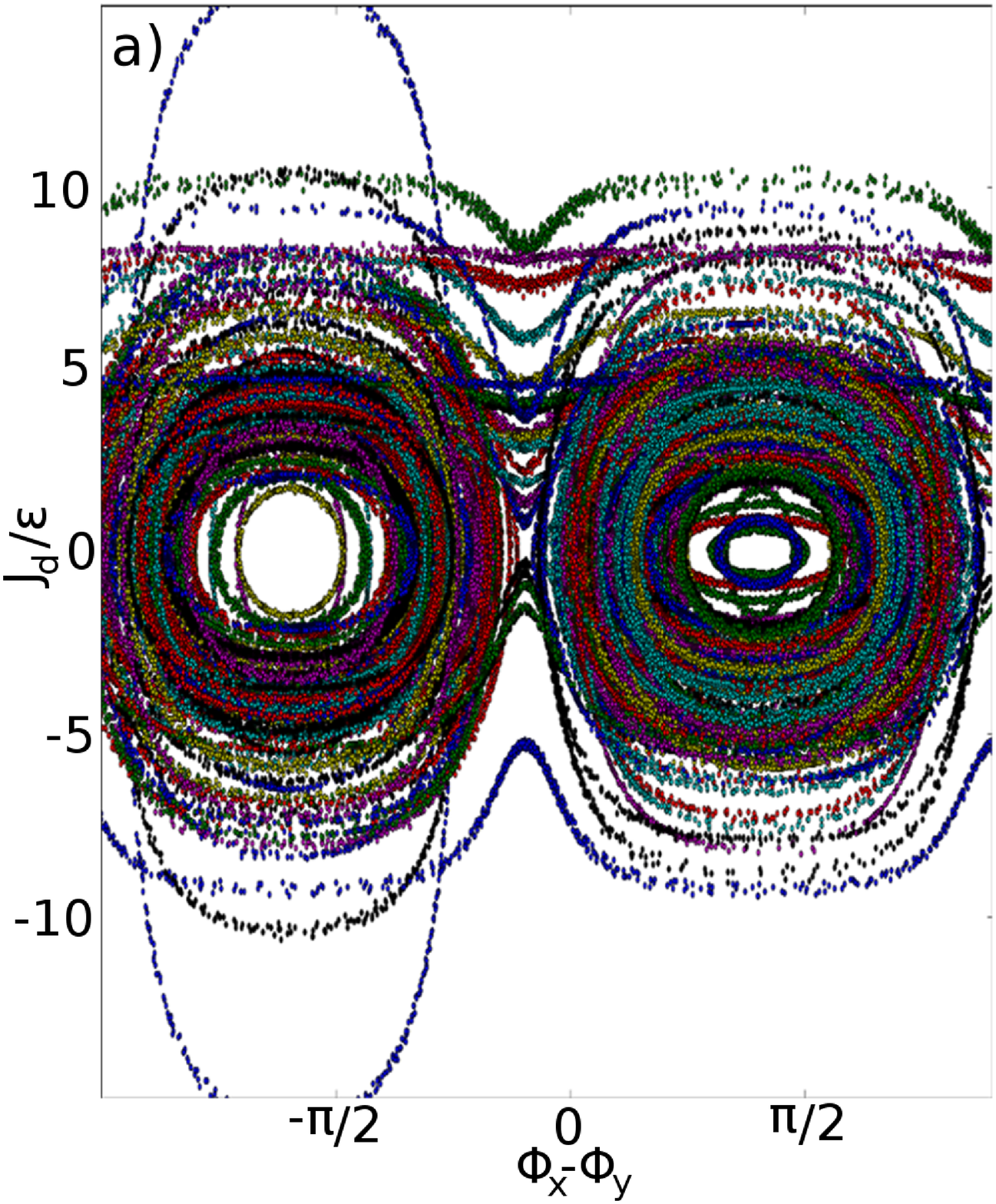}}
\subfloat{\includegraphics*[width=2.75in]{./spect_xjd.eps}}
\caption{Coupling resonance, $q=7.94$. (a) Poincare plots, $J_d$ versus $\Phi_x-\Phi_y$, for
randomly selected particles belonging to the  $0.5\%$ largest changing energy particles.
Most of these particles are trapped
in the resonance islands. (b) The horizontal tune density $\rho(Q_x)$ (black),
the $Q_t$ shifted tune density  $h(Q_x)$ (blue) and the one-tune-per-particle tune density $\rho_1(Q_x)$ (green)
with the corresponding $Q_t$ shifted tune density  $h_1(Q_x)$
for the $0.5\%$ largest changing energy  particles.
$h(Q_x)$ and $h_1(Q_x)$ are strongly peaked at the resonant mode tune $Q_x=\nu-Q_s$, showing 
mode-particle resonance via the $BJ_d\x$ term (see Eq.(\ref{eq:motionJd})).
}
\label{fig:poinc_fftjd}
\end{center}
\end{figure}

Most of the large changing energy  particles are  trapped  in resonance islands, as can be seen 
in Fig.~\ref{fig:poinc_fftjd}(a) where the Poincare plots, 
$J_{di}$ versus $\Phi_{xi}-\Phi_{yi}$ for $i \in S$, are shown. The phase coordinates are defined as
\begin{eqnarray}
\label{eq:phix}
\Phi_{xi}= \arctan \left(-\frac{\alpha_x x_i+ \beta_x x'_i}{x_i}\right),~~\Phi_{yi}= \arctan\left(-\frac{\alpha_y y_i+ \beta_y y'_i}{y_i} \right).
\end{eqnarray}
\noindent The Landau damping is, therefore, strongly influenced
by the $J_d$ oscillations  characterizing the coupling resonance trapped particles. 
The trapping frequency, $\omega_0 Q_{ti}$, of the LD-responsible particles is determined from the Fourier spectrum
of $J_{di}(s)$.

The  $Q_t$ satellites in the  particles' tune spectra contribute to the spectral weight at the mode coherent 
tune $\nu-Q_s$ by $\approx 20\% \sim 25\%$. This contribution favors the  conventional damping mechanism via the $A\x$ 
coupling (see Eq.(\ref{eq:motionJd})).
To estimate the $Q_t$ satellites' spectral weight we compare  the horizontal tune density
\begin{eqnarray}
\label{eq:rhox}
\rho_{x}(Q)=\sum_{i \in S} \rho_{xi}(Q)= \sum_{i \in S} |\tilde{x}_i(Q)|^2
\end{eqnarray}
and the one-tune-per-particle density 
\begin{eqnarray}
\label{eq:rho1x}
\rho_{1x}(Q)=\sum_{i \in S} \rho_{1xi}(Q)=\sum_{i \in S} \delta(Q-\Q_{xi}).
\end{eqnarray}
\noindent The sums in Eqs.(\ref{eq:rhox}) and (\ref{eq:rho1x}) are restricted only to the number of the selected particles with the largest energy change.  
$\Q_{xi}$ is the tune of the largest spectral peak in the Fourier spectrum $|\tilde{x}_i(Q)|$. 
Unlike $\rho_{x}$, where all spectral features are present, 
$\rho_{1x}$ assumes that every particle is characterized only by its main tune. The spectral weight difference between $\rho_{x}$ 
and $\rho_{1x}$ at  $\nu -Q_s $ measures the satellites contribution to the  $A\x$ damping mechanism. 
In Fig.~\ref{fig:poinc_fftjd}(b) $\rho_{x}$ and  $\rho_{1x}$ for the $0.5\%$ largest changing energy particles are shown.
Besides the peak at $\nu -Q_s$,  a broad spectral feature at smaller frequency, unfavorable 
to the $A\x$ damping mechanism, is observed in both $\rho_{x}$ and $\rho_{1x}$.

The other contribution of the $J_d$ oscillations to the damping is via the  $BJ_d\x$ term. 
The resonance condition is $Q_{xi} + Q_{ti} = \nu -Q_s $. We define  $h(Q)$  as the tune
density obtained by shifting each particle's horizontal tune by $Q_{ti}$, such
\begin{eqnarray}
\label{eq:h}
h(Q)&= \sum_{i \in S} h_i(Q)=\sum_{i \in S} \int  \rho_{Jdi}(Q')\rho_{xi}(Q-Q')dQ'\\ \nonumber
 &\approx \sum_{i \in S} \rho_{xi}(Q-Q_{ti}),
\end{eqnarray}
\noindent where $\rho_{Jdi}(Q)=|\tilde{J}_{di}(Q)|^2$ is the  Fourier spectrum of $J_{di}(s)$.
We define $h_1(Q)$ by replacing $\rho_{xi}$ with  $\rho_{1xi}$ in Eq.(\ref{eq:h}).
As shown in Fig.~\ref{fig:poinc_fftjd}(b), both $h(Q)$ and $h_1(Q)$ 
are strongly peaked at the coherent frequency $\nu -Q_s$ and  do not display 
the broad spectral feature seen in $\rho_{x}(Q)$ and $\rho_{1x}(Q)$ below $\nu-Q_s$.
In fact the particles with the tune forming the broad spectral feature of $\rho_{1x}(Q)$  have the main tune $\Q_{xi} \approx \nu-Q_s-Q_{ti}$, {\em i.e.},
the tune required for resonance with the $BJ_d\bar{x}$  coupling  term.
The broad feature seen in $h(Q)$ ($h_1(Q)$) spectrum above $\nu-Q_s$ corresponds to the spectral weight at $\nu-Q_s$ 
in the tune density $\rho_x(Q)$  ($\rho_{1x}(Q)$) when shifted with particle dependent $Q_{ti}$.

Compared to the off-resonance case, at coupling resonance the mode-particle coupling term $BJ_d\bar{x}$ 
allows a larger number of particles to participate to the damping process. The conventional
coupling does not favor particles with small longitudinal amplitudes, since they have a large 
tune shift which excludes them from the resonant exchange process
with the mode. However, this impediment is not so restrictive for the resonant exchange 
via the $BJ_d\bar{x}$ term, since  the trapping frequency $Q_t$ is proportional to the charge density,
thus also being  large for small longitudinal amplitude particles and partially compensating for the large tune shift.

\section{Discussions}
\label{sec:discuss}

\begin{figure}
\begin{center}
\includegraphics*[width=5.5in]{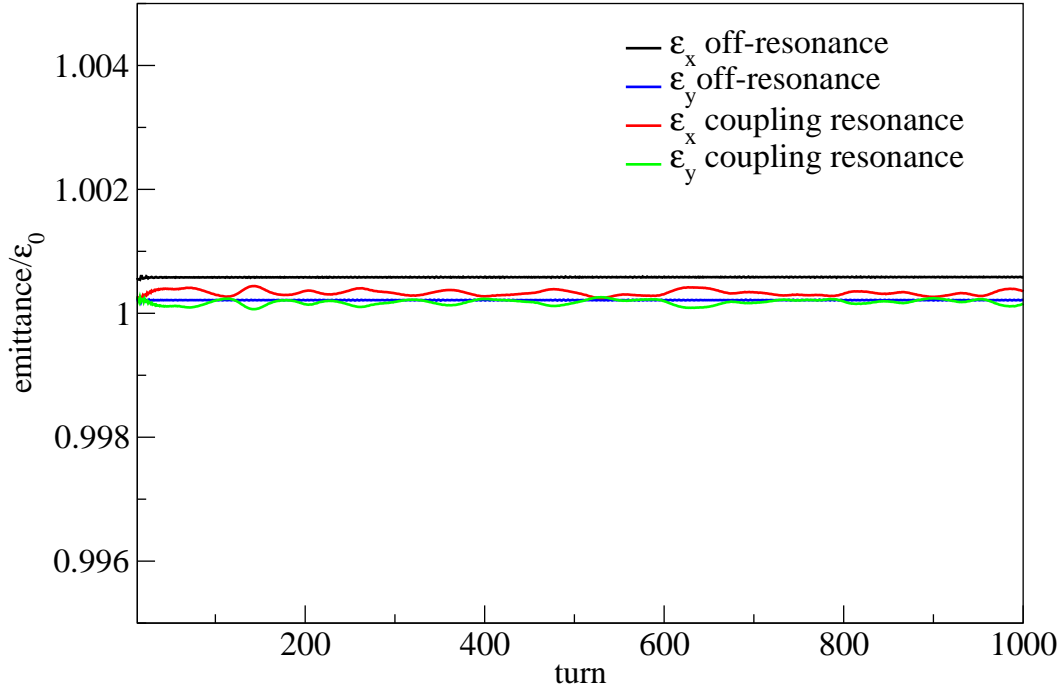}
\caption{The horizontal and the vertical emittances versus the turn number for the off-resonance and  the  coupling resonance 
cases for  $q=7.94$.
The emittances are nearly constant, the relative change being smaller than $5\times10^{-4}$.
}
\label{fig:m1_emitt}
\end{center}
\end{figure}

In our simulations we choose the initial amplitude excitation of the space charge mode to be small, 
of order of $10^{-3} \sigma_x$, where $\sigma_x$ is the beam horizontal rms size. As a consequence, 
the change in the beam shape and size are of the same order of magnitude. In a good approximation
the beam size is constant. One  may wonder about beam shape at the coupling resonance case, which, in general, is associated with 
the  Montague's emittance exchange between transverse planes~\cite{Montague}. 
The Montague's emittance exchange occurs when the  horizontal and the vertical emittances differs
significantly from each other.  In our case the initial beam distribution is chosen such that 
$\epsilon_x=\epsilon_y\equiv\epsilon_0$, and the emittance exchange is negligible. 
While at the particle level the coupling resonance is characterized by  amplitude exchange
between the transverse planes, the overall beam distribution changes very little.
In Fig.~\ref{fig:m1_emitt} we plot the transverse emittances versus the turn number,  both at the coupling resonance 
and in the off-resonance case. The relative change of the horizontal and the vertical emittances are smaller than $5 \times 10^{-4}$.

\begin{figure}
\begin{center}
\includegraphics*[width=5.5in]{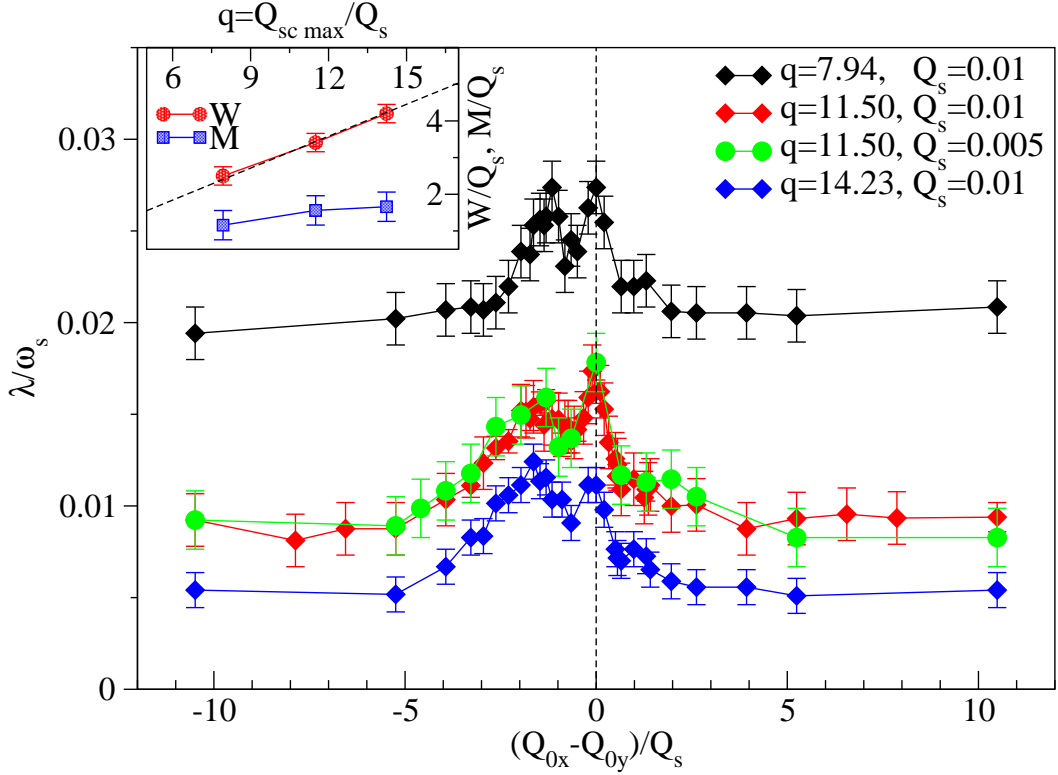}
\caption{Landau damping rate $\frac{\lambda}{\omega_s}$ of the horizontal first space charge mode 
versus $\Delta Q_0/Q_s=\left(Q_{0x}-Q_{0y}\right)/Q_s$
for different values of the space charge parameter $q$ and the synchrotron tune $Q_s$ where $\omega_s=\omega_0 Q_s$ is 
the synchrotron frequency.
The width of the enhanced damping region in the proximity of coupling resonance $W$,
defined at half maximum, scales linearly with the space charge tune shift,  $W \approx 0.3 Q_{sc~max}$, as shown in the inset.
The enhanced damping region is asymmetric with respect the to  $\Delta Q_0=0$, extending predominantly
on the negative side of $\Delta Q_0$. Two maxima of the damping rate can be noticed,  one at $\Delta Q_0=0$ 
and the other one at $\Delta Q_0 \approx -M$. 
}
\label{fig:m1_damping}
\end{center}
 \end{figure}

While our simulations show enhanced damping at the coupling resonance it is important to address
the damping behavior when moving away from this point.
We investigate the damping as a function of
the bare betatron tunes difference $\Delta Q_0=Q_{0x}-Q_{0y}$.  
In Fig.~\ref{fig:m1_damping} we plot the  Landau damping of the horizontal first space charge mode versus $\Delta Q_0/Q_s$
for different values of the space charge parameter and the synchrotron tune.
The width of the enhanced damping region scales linearly with 
the space charge tune shift, $W \approx 0.3 Q_{sc~max}$, where  $W$ is defined as the width at half maximum.
The enhanced damping region is asymmetric with respect the to  $\Delta Q_0=0$, extending predominantly ($\approx 80\%$)
on the negative side of $\Delta Q_0$.
We notice two maxima of the damping rate in the enhanced damping region. 
One is at coupling resonance $\Delta Q_0=0$ and the other at $-\Delta Q_0=M$. The value of $M$ increases slightly increasing $q$ as 
shown in the inset of Fig.~\ref{fig:m1_damping}. 
Our preliminary results indicate that the $-\Delta Q_0=M$ maximum and the asymmetry of the enhanced damping region are 
related to a different modulation of the mode-particle coupling. Those results will be addressed in detail in a future publication.

\section{Conclusions}
\label{sec:conclusions}

A novel Landau damping mechanism, driven by the modulation of the mode-particle coupling is introduced.
Numerical simulations using  Synergia with the DMD  method show the existence
of this mechanism in bunched beams in the proximity of coupling resonance.
The properties of the first space charge mode are calculated
for a Gaussian bunch propagating through an OFORODO lattice. The off-resonance
and the coupling resonance cases are compared. While the space charge mode's tune and shape
are nearly the same, the Landau damping is approximately a factor of $2$ larger
at coupling resonance in the strong space charge regime. In the off-resonance case the damping mechanism
can be understood within the conventional paradigm. The damping is caused by the resonant energy exchange
between the mode and the particles with an incoherent tune  equal to
the mode's tune shifted by $Q_s$.  At coupling resonance a large number
of particles are trapped around the stable points.
Their  transverse actions are oscillating
with a particle dependent trapping frequency $\omega_0 Q_t$. The spectral properties of 
the trapped particles with large energy exchange 
reveal that their tune is additionally shifted from the mode's coherent tune by $Q_t$,
supporting the  parametric Landau damping mechanism.

\section{Acknowledgments}
This work was performed at Fermilab, operated by Fermi Research Alliance, LLC under Contract No. DE-AC02-
07CH11359 with the United States Department of Energy.
Synergia development is partially supported through the ComPASS project, funded
through the Scientific Discovery through Advanced Computing program in the DOE Office of High Energy Physics.
An award of computer time was provided by the Innovative and Novel Computational Impact 
on Theory and Experiment (INCITE) program. This research used resources of 
the Argonne Leadership Computing Facility, 
which is a DOE Office of Science User Facility supported under Contract DE-AC02-06CH11357. 

\appendix
\section{Tune shift and mode-particle coupling}
\label{app:tsham}

The space charge induced tune shift and the coupling of the space charge collective modes
to the incoherent motion of the particles are dependent on the particles amplitudes. 
The interpretation of our numerical results is based on this assumption. 
For realistic beams it is difficult to derive analytical expressions for the tune shift and the mode-particle coupling.  
In this appendix we calculate these expressions
using a simply model. Despite the simplifying assumptions, the derivations provide insight on the 
dependence of the tune shift and the mode-particle coupling on the particles properties.

We consider a 2D model with  transverse degrees of freedom
\begin{eqnarray}
\label{eq:ham}
H=\frac{p_x^2}{2}+\frac{1}{2}\omega_{0x}^2 x^2+\frac{p_y^2}{2}+\frac{1}{2}\omega_{0y}^2 y^2+V(x ,y),
\end{eqnarray}
were  $V(x,y)$ is the space charge interaction part of the Hamiltonian.

\subsection{Off-resonance case}
\label{app:tsham_1}

We will introduce two different approaches for particle's frequency calculation.
In the first approach one writes the Hamiltonian in the canonical action-angle coordinates
\begin{eqnarray}
\label{eq:hamJ}
H= \omega_{0x} J_x+\omega_{0y} J_y +V(J_x,\ J_y, \Phi_x,\Phi_y),
\end{eqnarray}
\noindent where
\begin{eqnarray}
\label{eq:app_t2}
x=\sqrt{\frac{2 J_x}{\omega_{0x}}} \sin\Phi_x,~~
p_x=\sqrt{2 J_x\omega_{0x}} \cos \Phi_x.
\end{eqnarray}
\noindent Analogous transformations are used for $y$ and  $p_y$. Since the space charge potential contains only 
terms with even powers of $x$ and $y$, it can
be written as 
\begin{eqnarray}
\label{eq:Vspc0}
V(J_x,\ J_y, \Phi_x,\Phi_y)=V_0(J_x,J_y)+V_1(J_x,J_y,\cos 2 m \Phi_x, \cos 2 n \Phi_y),~m,n>0.
\end{eqnarray}
\noindent $V_1$ contains only high harmonics terms proportional to $\cos 2 m \Phi_x$
and/or  $\cos 2 n \Phi_y$.  Away from  resonances (see Appendix~\ref{app:tcr} for a discussion of the coupling resonance), 
the contribution of these terms  averages to zero in time. $V_1$ can be neglected in the first order of perturbation theory.
The  particle's horizontal frequency  is given by 
\begin{eqnarray}
\label{eq:tuneJ}
\omega_x=\dot{\Phi}_x=\frac{\partial (H_0+V_0)}{\partial J_x}=\omega_{0x}+ \frac{\partial V_0}{\partial J_x}.
\end{eqnarray}
The frequency shift is
\begin{eqnarray}
\label{eq:tuneshift}
\delta\omega_x(J_x,J_y)=\omega_{0x}-\omega_x=-\frac{\partial  V_0 }{\partial J_x}(J_x,J_y).
\end{eqnarray}
\noindent Analogous equations can be written for the vertical frequency.

The second approach for the particle's frequency calculation starts from the equation of motion and is based on the 
Lindstedt-Poincare perturbation theory~\cite{Lindstedt-Poincare} for nonlinear equations.
The horizontal displacement obeys
\begin{eqnarray}
\label{eq:eqmf}
\ddot{x}+ \omega_{0x}^2 x = F(x,y),
\end{eqnarray}
\noindent where 
\begin{eqnarray}
\label{eq:fv}
F(x,y)=-\frac{\partial V}{\partial x}(x,y)
\end{eqnarray}
\noindent is the space charge force considered to be a perturbation.
In the first order of perturbation theory, only the terms in $F(x,y)$ which 
oscillate with a frequency close to $\omega_x$  are relevant for the particle dynamics. To isolate these terms
we write $x$ and $y$ in the action-angle coordinates.
$F(x,y)$ contains terms with odd powers of $x$ and even powers of $y$ and it can be written as 
\begin{eqnarray}
\label{eq:Dspc0}
F(x,y)=F_0(J_x,J_y) \sin \Phi_x+F_1(J_x,J_y, \sin (2 m+1) \Phi_x, \cos 2 n \Phi_y)~m,n>0.
\end{eqnarray}
\noindent  $F_1$ can be neglected since
it contains only higher harmonics of $\omega_x=\dot{\Phi}_x$.  Eq.(\ref{eq:eqmf}) reduces to
\begin{eqnarray}
\label{eq:eqmf1}
\ddot{x}+ \left( \omega_{0x}^2  -F_0(J_x,J_y)\sqrt{\frac{\omega_{0x}}{2J_x}}  \right) x=0,
\end{eqnarray} 
and the horizontal tune shift is
\begin{eqnarray}
\label{eq:tuneshiftF}
\delta\omega_x(J_x,J_y)=F_0(J_x,J_y)\frac{1}{2 \sqrt{2 J_x\omega_{0x}}}.
\end{eqnarray}
\noindent Using Eqs.(\ref{eq:app_t2}) and (\ref{eq:fv}) one can check that Eq.(\ref{eq:tuneshift}) and  Eq.(\ref{eq:tuneshiftF}) 
agree with each other.

To estimate the mode-particle coupling we will use the equation of motion.
We consider a transverse mode which displaces the beam center infinitesimally by $\x$. 
Assuming the rigid-slice approximation,  the space charge potential in this case is
\begin{eqnarray}
\label{eq:Vmode}
V(x-\x,y)=V(x,y)-\frac{\partial V(x ,y)}{\partial x} \x.
\end{eqnarray}
The equation of motion reads
\begin{eqnarray}
\label{eq:f_cm}
\ddot{x}+ \omega_{0x}^2 x = F(x,y)-\frac{\partial F}{\partial x} (x,y) \x.
\end{eqnarray}
\noindent The  mode-particle coupling  is given by the last term in Eq.(\ref{eq:f_cm}).
In order to address the Landau damping mechanism, one has to investigate the parts in  the 
$\frac{\partial F}{\partial x} (x,y) \x$ term which are in resonance with the particle displacement $x$.
Let's assume in this section that the mode frequency  $\omega_c$ is close to
the particle's one $\omega_x$, as it happens for the conventional Landau mechanism. 
Since $\frac{\partial F}{\partial x} (x,y)$ contains only  even powers of $x$ and $y$ 
one can write
\begin{eqnarray}
\label{eq:df_terms}
\frac{\partial F}{\partial x}(x,y)=G_0(J_x,Jy)+ G_1 (J_x,J_y) \cos 2 \Phi_x+ G_2 (J_x,J_y,\cos 2 m \Phi_x, \cos 2n \Phi_y) \\
~m,n>0. \nonumber
\end{eqnarray}
\noindent Neglecting the  high harmonics,
the equation of motion 
can be written as 
\begin{eqnarray}
\label{eq:f_cm1}
\ddot{x}+ \left( \omega_{0x}-\delta \omega_x(J_x,J_y) \right) ^2  x \approx -G_0(J_x,Jy) \x +\frac{M}{2}  G_1 (J_x,J_y) \sin (2 \Phi_x-\omega_c t).
\end{eqnarray}
\noindent Here  $M$ is the mode amplitude, $\x = M \sin \omega_c t$, and $\delta \omega_x(J_x,J_y)$ is given by Eq.(\ref{eq:tuneshiftF}).
Since $\sin (2 \Phi_x-\omega_c t) \approx \sin (\omega_c t)$ when $\omega_c \approx \omega_x$ and the oscillations of $x$ and $\x$ are 
in phase~\cite{phase_footnote}, 
one can write the equation of motion as 
\begin{eqnarray}
\label{eq:cmk}
\ddot{x}+ \left( \omega_{0x}-\delta \omega_x(J_x,J_y) \right) ^2 x = -K(J_x,J_y) \x,
\end{eqnarray}
\noindent with the mode-particle coupling given by
 \begin{eqnarray}
\label{eq:kjxjy}
K(J_x,J_y)=G_0(J_x,J_y)- \frac{1}{2} G_1 (J_x,J_y).
\end{eqnarray}

We will end this section with an example.
Let's consider  a  space charge  potential 
\begin{eqnarray}
\label{eq:Vx4}
V(x)=\alpha x^4.
\end{eqnarray}
\noindent This is not a realistic potential but can be viewed as a term in the Taylor's expansion of the real potential.
The space charge force is
\begin{eqnarray}
\label{eq:Fx4}
F(x)=-4 \alpha x^3= -4 \alpha \frac{2J_x}{\omega_{0x}}\sqrt{\frac{2J_x}{\omega_{0x}}} \sin^3 \Phi
=-4 \alpha \frac{2J_x}{\omega_{0x}}\sqrt{\frac{2J_x}{\omega_{0x}}} \left( \frac{3}{4} \sin \Phi_x -\frac{1}{4} \sin 3 \Phi_x \right). 
\end{eqnarray}
Therefore 
\begin{eqnarray}
\label{eq:F0x4}
F_0(J_x)= -\alpha \frac{6J_x}{\omega_{0x}}\sqrt{\frac{2J_x}{\omega_{0x}}},
\end{eqnarray}
\noindent and according to Eq.(\ref{eq:tuneshiftF}) the tune shift is
\begin{eqnarray}
\label{eq:F0x4_ts}
\delta\omega_x(J_x)=-3 \alpha \frac{J_x}{\omega^2_{0x}}.
\end{eqnarray}

For the mode-particle coupling we have
\begin{eqnarray}
\label{eq:mpcx4}
\frac{\partial F}{\partial x}(x,y)=- 12 \alpha x^2=-12 \alpha \frac{ J_x}{\omega_{0x}} (1-\cos 2 \Phi_x),
\end{eqnarray}
\noindent implying
\begin{eqnarray}
\label{eq:Kx4}
K(J_x)= - 18 \alpha \frac{J_x}{\omega_{0x}}= 3 \times  2 \omega_{0x} \delta\omega_x(J_x).
\end{eqnarray}
The factor of $3$ in front of the rhs of the Eq.(\ref{eq:Kx4}) arises from the derivative of $F(x,y)$ with respect to $x$.
A  space charge force term  $F(x) \propto x^{2n+1}$, implies $K(J_x)= (2n+1) \times  2 \omega_{0x} \delta\omega_x(J_x)$.
In the linear regime where $F(x) \propto x$, valid for small amplitude  oscillations,
$K(J_x)=2 \omega_{0x} \delta\omega_x(J_x)$.

Note that the  equation  Eq.(\ref{eq:motion}) proposed to describe the particles motion in Section~\ref{sec:scm}, is a generalization
of Eq.(\ref{eq:cmk}) to include longitudinal  degrees of freedom.

\subsection{Transverse coupling resonance for round beams}
\label{app:tcr}

The resonance effects on beam particles dynamics are treated extensively in the literature.
However the question of how the resonance effects can influence
the mode-particle coupling and implicitly the Landau damping mechanism has not been addressed.
In this section, using  a simplified space charge potential, we will discuss the mode-particle coupling in the proximity
of transverse coupling resonance given by  $2 \omega_x - 2  \omega_y=0$.

To illustrate the calculation of tune shift and mode-particle coupling 
near coupling resonance we consider an example where the space charge potential is rotationally symmetric  and given by 
\begin{eqnarray}
\label{eq:app_MH}
V(x,y)=\alpha \left(x^2+y^2 \right)^2.
\end{eqnarray}

In the proximity of the coupling resonance it is easier to calculate the particles dynamics with a Hamiltonian formalism
than by using the equation of motion.  For the calculation of  the frequency shift we
transform the Hamiltonian to action-angle coordinates 
\begin{eqnarray}
\label{eq:app_H11}
H=\omega_{0x} J_x+\omega_{0y} J_y+ \frac{3 \alpha}{2}\left( \frac{J_x^2}{\omega_{0x}^2} + \frac{J_y^2}{\omega_{0y}^2}\right)+
2 \alpha \frac{J_x J_y}{\omega_{0x} \omega_{0y}}\left(1+\frac{1}{2}\cos (2\Phi_x -2\Phi_y) \right).
\end{eqnarray}
\noindent We neglect the high harmonics terms, as discussed in Sec.~\ref{app:tsham_1}. Assuming
proximity of the  coupling resonance  the term proportional to $\cos (2\Phi_x -2\Phi_y)$ is retained.

\noindent We proceed with a  canonical transformation using the generating functional 
\begin{eqnarray}
\label{eq:app_F2}
 F(\Phi_x,\Phi_y,I_x,I_y)= \left(\Phi_x- \bar{\omega}t\right) I_x+\left(\Phi_y- \bar{\omega}t\right) I_y,
\end{eqnarray}
\noindent with $\bar{\omega}=\frac{1}{2}(\omega_{0x}+\omega_{0y})$. It leads to the Hamiltonian
\begin{eqnarray}
\label{eq:app_t12}
\tilde{H}= \Delta \omega (-I_x+I_y)+\frac{3 \alpha}{2}\left( \frac{I_x^2}{\omega_{0x}^2} + \frac{I_y^2}{\omega_{0y}^2}\right)
+2 \alpha \frac{I_x I_y}{\omega_{0x} \omega_{0y}}\left(1+\frac{1}{2}\cos (2\varphi_x -2\varphi_y) \right),
\end{eqnarray}
\noindent where 

\begin{eqnarray}
\label{eq:app_t11}
\Delta \omega =\frac{1}{2}(\omega_{0y}-\omega_{0x})\\
\varphi_x=\Phi_x-\bar{\omega}t,~~
J_x=I_x\\
\varphi_y=\Phi_y-\bar{\omega}t,~~
J_y=I_y\\
\tilde{H}=H-\bar{\omega}(I_x+I_y).
\end{eqnarray}

A subsequent  canonical transformation using the generating functional
\begin{eqnarray}
\label{eq:app_F3}
 F(\varphi_x,\varphi_y,J_s,J_d)= \frac{1}{2} J_s (\varphi_x+\varphi_y)+\frac{1}{2} J_d (\varphi_x-\varphi_y),
\end{eqnarray}
\noindent leads to the Hamiltonian

\begin{eqnarray}
\label{eq:app_Hfinal}
H=-\Delta \omega J_d+\alpha \left[a_s J_s^2 + a_d J_d^2 + b J_s J_d +k (J_s^2-J_d^2 )\cos (4 \varphi_d) \right]
\end{eqnarray}

\noindent where  the canonical variables are
\begin{eqnarray}
\label{eq:app_t13}
\varphi_s=\frac{1}{2}  (\varphi_x+\varphi_y),~~
J_s=I_x+I_y\\
\varphi_d= \frac{1}{2}  (\varphi_x-\varphi_y),~~
J_d=I_x-I_y\\
\end{eqnarray}
\noindent and

\begin{eqnarray}
\label{eq:app_H_coeff}
a_s=&\frac{3}{8}\left(\frac{1}{\omega_{0x}^2}+ \frac{1}{\omega_{0y}^2}+\frac{4}{3 \omega_{0x} \omega_{0y}}\right)\\
a_d=&\frac{3}{8}\left(\frac{1}{\omega_{0x}^2}+ \frac{1}{\omega_{0y}^2}-\frac{4}{3 \omega_{0x} \omega_{0y}}\right)\\
b=&\frac{3}{4}\left(\frac{1}{\omega_{0x}^2}- \frac{1}{\omega_{0y}^2}\right)\\
k=&\frac{1}{4} \frac{1}{\omega_{0x} \omega_{0y}}.
\end{eqnarray}

Since the Hamiltonian is independent on $\varphi_s$, it follows that  $J_s$ is a constant of motion.

The stable point $(\varphi^*_d, J^*_d)$  is given by the the equations 
$\dot{J_d}=0$, $\dot{\varphi_d}=0$ which lead to
\begin{eqnarray}
\label{eq:app_stable}
\cos (4 \varphi^*_d)=-1,~~
J^*_d=\frac{\Delta \omega \bar{\omega}^2}{\alpha},
\end{eqnarray}
\noindent up to the first order in $\Delta \omega$.
The particle dynamics around the stable point is described by the effective Hamiltonian

\begin{eqnarray}
\label{eq:app_Hstable}
H_{eff}=\frac{\alpha}{2 \bar{\omega}^2} \delta J_d^2+2 \frac{\alpha}{\bar{\omega}^2}(J_s^2-{J^*}_d^2 )\delta \varphi_d^2
\end{eqnarray}
\noindent where $\delta J_d=J_d-J^*_d$, $\delta \varphi_d=\varphi_d-\varphi^*_d$.
The oscillation frequency of $J_d$ in the proximity of the coupling resonance stable point is therefore
\begin{eqnarray}
\label{eq:app_wtr}
\omega_{t}=\frac{2 \alpha}{\bar{\omega}^2}\sqrt{(J_s^2-{J^*}_d^2 )}.
\end{eqnarray}

Considering that
\begin{eqnarray}
\label{eq:app_tune_shr}
\dot{\varphi}_s=2 \alpha (a_s+k \cos(4 \varphi_d))J_s=2 \frac{\alpha}{\bar{\omega}^2} J_s\\
\dot{\varphi}_d= \frac{\alpha} {\bar{\omega}^2} \delta J_d
\end{eqnarray}
\noindent  the particle tunes are
\begin{eqnarray}
\label{eq:app_xtune_shr}
\omega_x \approx \dot{\Phi}_x = \dot{\varphi}_s+\dot{\varphi}_d+ \bar{\omega}=2 \frac{\alpha}{\bar{\omega}^2}  J_s + \bar{\omega} + \frac{\alpha}{\bar{\omega}^2} (J_d -J_d^{*}) \\
\label{eq:app_ytune_shr}
\omega_y \approx \dot{\Phi}_y  =\dot{\varphi}_s-\dot{\varphi}_d+ \bar{\omega}=2 \frac{\alpha}{\bar{\omega}^2}  J_s + \bar{\omega} - \frac{\alpha}{\bar{\omega}^2} (J_d -J_d^{*}).
\end{eqnarray}
\noindent Note that in the frequency space the main spectral feature of the trapped particle resides on the line $\omega_x=\omega_y$.
The oscillation of $J_d$  yields satellite features spaced by harmonics of the trapping frequency $\omega_t$.

The frequency shift is given by
\begin{eqnarray}
\label{eq:app_xytune_shift}
\delta \omega_x (J_s,J_d) =\omega_{0x}-\omega_{x} =\Delta \omega + 2 \frac{\alpha}{\bar{\omega}^2}  J_s + \frac{\alpha}{\bar{\omega}^2} (J_d -J_d^{*}) \\
\delta\omega_y (J_s,J_d) = \omega_{0y}-\omega_{y} =-\Delta \omega + 2 \frac{\alpha}{\bar{\omega}^2}  J_s - \frac{\alpha}{\bar{\omega}^2} (J_d -J_d^{*}).
\end{eqnarray}

In the presence of a transverse mode, in the rigid-slice approximation, the space charge potential is
\begin{eqnarray}
\label{eq:app_VPM}
V(x,y)=\alpha \left((x-\x)^2+y^2 \right)^2.
\end{eqnarray}
\noindent The calculation of the mode-particle coupling reduces to analyzing the force acting on the particle.
The equation of motion reads,
\begin{eqnarray}
\label{eq:cr_eqm}
\ddot x + \omega^2_{0x} x= -4 \alpha x (x^2+y^2)+ 12 \alpha x^2 \x +4 \alpha y^2 \x.
\end{eqnarray} 
\noindent The
displacement $\x$ is small and does not affect the incoherent particle motions in the first order approximation.
Eq.(\ref{eq:cr_eqm}) yields  the mode-particle coupling written in action-angle coordinates as
\begin{eqnarray}
\label{eq:app_pmc2}
-K \x& =& \left( 12 \alpha x^2+ 4 \alpha y^2 \right) \x\\ \nonumber
&=&\left( 12 \alpha \frac{(J_s+J_d)}{\omega_{0x}} \sin^2 \Phi_x + 4 \alpha \frac{(J_s-J_d)}{\omega_{0y}} \sin^2 \Phi_y  \right) \x \\ \nonumber
&=&2 \alpha J_s \left( \frac{3}{\omega_{0x}} (1-\cos 2 \Phi_x) + \frac{1}{\omega_{0y}} (1-\cos 2 \Phi_y) \right) \x \\ \nonumber
&&+2 \alpha J_d \left( \frac{3}{\omega_{0x}} (1-\cos 2 \Phi_x) - \frac{1}{\omega_{0y}} (1-\cos 2 \Phi_y) \right) \x.
\end{eqnarray}

Because $J_s$ is constant, the first term in the rhs of Eq.(\ref{eq:app_pmc2}) is resonant with $x$ when 
$\omega_x \approx \omega_y \approx \omega_c$, as expected by 
a conventional Landau damping mechanism. A similar reasoning with the one discussed in  Appendix~\ref{app:tsham_1}
can be used to calculate the mode-particle coupling. However, when analyzing the terms resonant with 
$x$ one has to take into account that, according to Eq.(\ref{eq:app_stable}),  
the trapped particles phases satisfy $2 \Phi_x-2 \Phi_y=(2k+1)\pi$. Doing that, one gets for the coupling term proportional to $J_s$,
\begin{eqnarray}
\label{eq:app_jscp}
K_1\x=- J_s \left(\frac{9 \alpha  }{\omega_{0x}} +\frac{\alpha  }{\omega_{0y}}\right) \x \approx -\frac{10 \alpha J_s }{\omega_{0x}} \x.
\end{eqnarray}

On the other hand, since $J_d$ is oscillating with the frequency $\omega_t$, the term in rhs of Eq.(\ref{eq:app_pmc2}) proportional to $J_d$ 
is resonant with $x$  when $\omega_x \approx \omega_y \approx \omega_c-\omega_t $. 
Assuming  that $J_d=|J_d|\cos \omega_t t$, $\x=M \sin \omega_c t$ and $2 \Phi_x-2 \Phi_y=(2k+1)\pi$,
the mode-particle coupling proportional to $J_d$ can be written as
\begin{eqnarray}
\label{eq:app_jdcp}
K_2\x=-J_d\left( \frac{15 \alpha }{2\omega_{0x}} -\frac{3 \alpha }{2\omega_{0y}} \right) \x \approx  -\frac{6 \alpha J_d }{\omega_{0x}} \x.
\end{eqnarray}

The equation of motion in the proximity of the coupling resonance for our simplified Hamiltonian is 
\begin{eqnarray}
\label{eq:app_gpmc}
\ddot x + \left(\omega_{0x}-\delta \omega_x(J_s,J_d)\right)^2 x = -(K_1+K_2)\x=\frac{10 \alpha }{\omega_{0x}} J_s \x + \frac{6 \alpha}{\omega_{0x}} J_d \x.
\end{eqnarray}

The equation Eq.(\ref{eq:motionJd})  proposed to described the motion of particles in the proximity of coupling resonance 
in Section~\ref{sec:scm} is a generalization of Eq.(\ref{eq:app_gpmc})
to include higher order terms in the space charge potential and, as well,   
longitudinal degrees of freedom.


\begin{thebibliography}{99}
\bibitem{phase_footnote} For the Landau damping mechanism only the part of the solution in phase with 
the driving force is relevant, see chapter 5 in ~\cite{Chao_book}.
\end{thebibliography}
\end{document}